\documentstyle[preprint,aps,pre]{revtex}

\begin{document}

\title{Disordering effects of colour in a system of coupled
Brownian motors: phase diagram and anomalous-to-normal hysteresis
transition}

\author{H. S. Wio $^1$\thanks{Member of CONICET, Argentina; E-mail:
wio@cab.cnea.gov.ar}, S.  E.  Mangioni$^2$ and R.  R.  Deza$^2$}
\address{1) Grupo de F\'{\i}sica Estad\'{\i}stica
\thanks{http://www.cab.cnea.gov.ar/CAB/invbasica/FisEstad/estadis.htm} \\
Centro At\'omico Bariloche (CNEA) and Instituto Balseiro
(CNEA and UNCuyo)\\ 8400 San Carlos de Bariloche, Argentina \\
2) Departamento de F\'{\i}sica, FCEyN, Universidad Nacional de Mar
delPlata\\ De\'an Funes 3350, 7600 Mar del Plata, Argentina.}

\maketitle

\begin{abstract}
A system of periodically coupled nonlinear phase oscillators\textemdash
submitted to both additive and multiplicative white noises\textemdash
has been recently shown to exhibit ratchetlike transport, negative
zero-bias conductance, and anomalous hysteresis.  These features stem
from the {\em asymmetry\/} of the stationary probability distribution
function, arising through a noise-induced nonequilibrium phase
transition which is {\em reentrant\/} as a function of the
multiplicative noise intensity.  Using an explicit mean-field
approximation we analyze the effect of the multiplicative noises being
coloured, finding a contraction of the ordered phase (and a reentrance
as a function of the coupling) on one hand, and a shift of the
transition from anomalous to normal hysteresis inside this phase on the
other.
\end{abstract}

\section{Introduction}\label{sec.1}
One of the distinguishing features of the last decade has been a
spectacular advancement of knowledge in the field of
{\em nanotechnology\/}.  Within a subfield that we might call
``nanomechanics'', the topic of {\em noise-induced transport\/}
\cite{[1]} \textemdash which concerns the mechanisms whereby one can
extract useful work out of (nonequilibrium) fluctuations\textemdash
has increasingly captivated researchers.

In the early works it seemed to be a requisite for the operation of
microscopic rectifying devices (usually known as ``molecular motors'',
``Brownian motors'', or ``ratchets'') that\textemdash besides a
built-in ratchetlike bias\textemdash the fluctuations be correlated
\cite{[2]}.  That requirement was relaxed when ``pulsating'' ratchets,
in which the rectifying effect arises from the random {\em switching\/}
between uncorrelated noise sources, were discovered \cite{[3]}.  A
recent twist has been to relax also the requirement of a built-in
bias: a system of periodically coupled nonlinear phase oscillators in
a symmetric ``pulsating'' environment has been shown to undergo a
noise-induced nonequilibrium phase transition, wherein the spontaneous
symmetry breakdown of the stationary probability distribution function
(SPDF) gives rise to an {\em effective\/} ratchetlike potential.  Some
of the striking consequences of this fact are the appearance of {\em
negative\/} (absolute) zero-bias conductance in the disordered phase,
but near the phase-transition line (for small values of the bias force
$F$, the particle current $\langle\dot X\rangle$ opposes $F$), and
{\em anomalous hysteresis\/} in the strong-coupling region of the
ordered phase (the $\langle\dot X\rangle$ vs $F$ cycle runs {\em
clockwise\/}, as opposed for instance to the $B$ vs $H$ cycle of a
ferromagnet) \cite{[4]}.

Exploiting our previous experience \cite{[5]}, in Ref.\ \cite{[6]} we
addressed the model using an explicit mean-field approach (see e.g.\
Ref.\ \cite{[7]}).  We undertook a thorough exploration of the ordered
phase\textemdash including the characterization of its subregions and
the transition from anomalous to normal hysteresis in the behaviour of
$\langle\dot X\rangle$ as a function of $F$\textemdash and found a
close relationship between the shape of the SPDF and the number of
``homogeneous'' mean-field solutions on one hand, and the character of
the hysteresis loop on the other.

As discussed in Ref.\ \cite{[5]}, the multiplicative noises are
expected to exhibit some degree of time-correlation or ``colour''.
Hence in this work, and as a natural continuation to Ref.\ \cite{[6]},
we study (in mean-field approximation) the consequences of a {\em
finite\/} correlation time $\tau$ of the multiplicative noises in the
model of Refs.\ \cite{[4],[6]}.  In the following sections we briefly
describe the model and its mean-field treatment, and discuss our
numerical results regarding the influence of colour on the phase
diagram and on the transition from anomalous to normal hysteresis
inside the ordered phase.

\section{The model and its mean-field analysis}\label{sec.2}
\textbf{The model:} Let us consider the following set of $N$ {\em
globally\/} coupled stochastic equations of motion, in the overdamped
regime:
\begin{equation}
    \dot X_i=-\frac{\partial U_i}{\partial X_i}+\sqrt{2T}\,
    \xi_i(t)-\frac{1}{N}\sum_{j=1}^N K(X_i-X_j).\label{eq:1}
\end{equation}
The stochastic variables $X_i(t)$ are phaselike [$-L/2\leq X_i(t)\leq
L/2$, where $L$ is the period of the $U_i$] and the equation is meant
to be interpreted in the sense of Stratonovich.  The second term in
Eq.\ (\ref{eq:1}) models, as usual, the effect of thermal fluctuations
[the $\xi_i(t)$ are {\em additive\/} Gaussian white noises with zero
mean and variance one: $\langle\xi_i(t)\rangle=0\mbox{ ,
}\langle\xi_i(t)\xi_j(t')\rangle= \delta_{ij}\delta(t-t')$, and $T$
represents the temperature of the environment].

The ``pulsating'' potentials $U_i(x,t)$ are one of the key ingredients
in the model \cite{[3]}.  Including a ``load force'' $F$ as a tool for
the analysis of the noise-induced ratchet effect, their form is
\begin{equation}
    U_i(x,t)=V(x)+W(x)\sqrt{2Q}\,\eta_i(t)-Fx,\label{eq:2}
\end{equation}
namely they consist of a static part $V(x)$ and a fluctuating one:
Gaussian noises $\eta_i(t)$ with zero mean couple {\em
multiplicatively\/} (with intensity $Q$) through a function $W(X_i)$.
Whereas in Refs.\ \cite{[4],[6]} the $\eta_i(t)$ were taken as white
with variance one, we now regard them instead as Ornstein-Uhlenbeck,
with self-correlation time $\tau$:
$\langle\eta_i(t)\eta_j(t')\rangle=\delta_{ij}(\sigma^2/2\tau)
\exp(-|t-t'|/\tau)$.

Besides being {\em periodic\/}, $V(x)$ and $W(x)$ are assumed to be
{\em symmetric\/} $V(-x)=V(x)$ and $W(-x)=W(x)$ (there is no
{\em built-in\/} ratchet effect).  In Refs.\ \cite{[4],[6]} it is
$V(x)=W(x)=-\cos x-A\cos{2x}$, hence $L=2\pi$.  With the choice $A>0$,
the direction of the particle current $\langle\dot X\rangle$ turns out
to be {\em opposite\/} to that of symmetry breaking in the SPDF
$P^{st}(x)$: it is this effect which leads in turn to such oddities as
{\em negative zero-bias conductance\/} and {\em anomalous hysteresis\/}
\cite{[4]}.  The interaction force $K(x-y)=-K(y-x)$ between oscillators
is a periodic function of $x-y$ (also with period $L$) and in
Refs.\ \cite{[4],[6]} is chosen as $K(x-y)=K_0\sin(x-y)$, with $K_0>0$.
We shall fix $T=2.0$ and $A=0.15$ as in Refs.\ \cite{[4],[7]}, so the
important parameters in the model are $K_0$ and $Q$.  The model just
set up can be visualized (at least for $A\to 0$) as a set of
overdamped pendula (only their phases matter, not their locations)
interacting with one another through a force proportional to the
$\sin$ of their phase difference (this force is always attractive in
the reduced interval $-\pi\le x-y\le\pi$).

\textbf{Mean-field approximation (MFA)} With the above choice, the
interparticle interaction term in Eq.\ (\ref{eq:1}) can be cast in the
form
\begin{equation}
    \frac{1}{N}\sum_{j=1}^N K(X_i-X_j)=
    K_0\left[C_i(t)\sin X_i-S_i(t)\cos X_i\right].\label{eq:3}
\end{equation}
For $N\to\infty$ we may approximate Eq.\ (\ref{eq:3}) \`a la
Curie-Weiss, replacing $C_i(t)\equiv N^{-1}\sum_j\cos x_j(t)$ by
$C_m\equiv\langle\cos x_j\rangle$ and $S_i(t)\equiv N^{-1}
\sum_j\sin x_j(t)$ by $S_m\equiv\langle\sin x_j\rangle$, to be
determined as usual by self-consistency.  This decouples the
system of stochastic differential equations (SDE) in Eq.\
(\ref{eq:1}), which reduces to essentially one SDE for the single
stochastic process $X(t)$. {\em If\/} $X(t)$ happens to be
Markovian, then it is a straightforward matter to write up an
associated Fokker-Planck equation (FPE) whose stationary solution
is the SPDF, from which all the transport properties can be
readily obtained.  As discussed in Ref.\ \cite{[3]}, the Gaussian
character of the $\eta_i(t)$ allows in the $\tau=0$ case to
consider them as being coupled through an effective function
$S(X_i)\equiv\sqrt{2}[T+Q(W')^2]^{1/2}$.  Hence in the white-noise
case, the SDE obtained after performing the MFA is
\begin{equation}
    \dot X=R(X)+S(X)\eta(t),\label{eq:4}
\end{equation}
with $R(x)=-V'(x)+F-K_0[C_m\sin x-S_m\cos x]$.  The associated
FPE is
\begin{equation}
    \partial_t P(x,t)=\partial_x\{-[R(x)+\frac{1}{2}S(x)\,S'(x)]P(x,t)
    \}+\frac{1}{2}\partial_{xx}[S^2(x)P(x,t)]\label{eq:5}
\end{equation}
and its normalised stationary solution {\em with periodic boundary
conditions\/} and current density $J\ne 0$ is \cite{[3],[4]}
\begin{equation}
    P^{st}(x)=\frac{e^{-\phi(x)}\,H(x)}{{\cal N}\,S(x)},\label{eq:6}
\end{equation}
where $\phi(x)=-2\int_0^x dy\,[R(y)/S^2(y)]$,
$H(x)=\int_{x}^{x+L}dy\,S(y)^{-1}\, \exp[\phi(y)]$, and ${\cal
N}=\int_{-L/2}^{L/2}dx\,P^{st}(x)$.  The positivity of $S(x)$ and
the exponentials implies that of $H(x)$ and hence that of
$P^{st}(x)$ and ${\cal N}$, as it should be.

\textbf{The particle current:} The appearance of a ratchet effect
amounts to the existence of a nonvanishing drift term $\langle\dot
X\rangle$ in the stationary state, in the absence of any forcing
($F=0$); in other words, the pendula become rotators in an average
sense.  As it was shown in Ref.\ \cite{[4]}, the cause of this
spontaneous particle current is the noise-induced asymmetry in
$P^{st}(x)$. Being the current density
\begin{equation}
    J=[1-e^{\phi(L)}]/2{\cal N},\label{eq:7}
\end{equation}
the sign of $J$ is that of $1-e^{\phi(L)}$. The ``holonomy'' condition
$e^{\phi(L)}=1$ implies $J=0$ and $H(x)=\mathrm{const}$.  As shown in
Ref.\ \cite{[6]} it is
\begin{equation}
    \langle\dot X\rangle=
    \int_{-L/2}^{L/2}dx\left[R(x)+\frac{1}{2}S(x)S'(x)\right]
    P^{st}(x,C_m,S_m),\label{eq:8}
\end{equation}
with the result
\begin{equation}
    \langle\dot X\rangle=J\,L=\left\{\frac{1-e^{\phi(L)}}{2{\cal N}}
    \right\}L,\label{eq:9}
\end{equation}
hence $\langle\dot X\rangle$ has the sign of $J$ and can be also
regarded as an order parameter.

Equation (\ref{eq:7}) is a self-consistency relation since both
${\cal N}$ and $\phi(L)$ carry information on the shape of $P^{st}(x)$
(in the latter case through $C_m$ and $S_m$).  A nonzero $J$ is always
associated with a symmetry breakdown in $P^{st}(x)$ (namely,
$P^{st}(-x)\neq P^{st}(x)$).  This may be either {\em spontaneous\/}
(our main concern here) or {\em induced\/} by a nonzero $F$.

\textbf{Effective Markovian approximation for coloured noise:} The
results in Eqs.\ (\ref{eq:6}) and (\ref{eq:9}) rely on the fact that we
have been able to write the FPE Eq.\ (\ref{eq:5}).  When the $\eta_i(t)$
in Eq.\ (\ref{eq:2}) are coloured, the process $X(t)$ in Eq.\
(\ref{eq:4}) is in principle not Markovian and for non-Markovian
processes, a FPE can at most result from some (non-systematic)
approximation, like the truncation of some short correlation-time
expansion.  Fortunately, a consistent Markovian approximation (called
``unified coloured-noise approximation'' or UCNA) can be performed under
certain conditions.  By resorting to it one can obtain expressions for
$R(x)$ and $S(x)$ in Eq.\ (\ref{eq:5}) which account for the effect of
$\tau$.  Their functional forms will be published elsewhere \cite{[8]}.

\textbf{The self-consistency equations:} The stationary probability
distribution $P^{st}(x)$ also depends on $S_m$ and $C_m$, since $R(x)$
contains these parameters.  Their values arise from requiring
self-consistency, which amounts to solving the following system of
nonlinear integral equations:
\begin{eqnarray}
    F_{cm}&=&C_m,\mbox{   with  }F_{cm}\equiv\langle\cos x\rangle=
    \int_{-L/2}^{L/2}dx\,\cos x\,P^{st}(x,C_m,S_m),\label{eq:10}\\
    F_{sm}&=&S_m,\mbox{   with  }F_{sm}\equiv\langle\sin x\rangle=
    \int_{-L/2}^{L/2}dx\,\sin x\,P^{st}(x,C_m,S_m).\label{eq:11}
\end{eqnarray}
These equations give $C_m$ and $S_m$ for each set of the parameters
($Q$, $K_0$) that define the state of the system, assumed $T$, $A$ and
$F$ fixed.

For $F=0$, the choice $S_m=0$ makes $R(x)$ an odd function of $x$;
this in turn makes $\phi(x)$ even, and then the periodicity of
$P^{st}(x)$ in Eq.\ (\ref{eq:6}) [in the form$P^{st}(-x)=P^{st}(-x-L)$]
implies that the stationary probability distribution is also an even
function of $x$.  So the problem of self-consistency reduces to the
{\em numerical\/} search of solutions to Eq.\ (\ref{eq:10}), with
$S_m=0$. Although plausibility arguments, detailed in Ref.\ \cite{[6]},
allow to have an intuition on the existence of some solutions to this
integral equation (and their stability) in this {\em symmetric\/} case,
the stability of the {\em true\/} solutions must be explicitly checked.
Since $\cos x$ in Eq.\ (\ref{eq:10}) is an even function of $x$, it
suffices to use the Curie-Weiss one-parameter criterion, namely to
check whether the slope at $S_m$ of the integral in Eq.\ (\ref{eq:11})
is less or greater than one.  As a complementary check, a small-$x$
expansion of $\phi(x)$ \cite{[6]} confirms that $P^{st}(x)$ is indeed
Gaussian at $x=0$.  For {\em small\/} $F\neq 0$, $P^{st}(x)$ gets
multiplied (in this approximation) by $\exp[Fx/T]\,(\cong 1+Fx/T)$
which leads to a nonzero value of $S_m=kF$, with $k>0$.  By the
mechanism discussed in Ref.\ \cite{[6]}, for large enough $Q$ it is
$\phi(L)>0$ and by Eq.\ (\ref{eq:7}) it is $J<0$.  This effect
manifests itself in a {\em negative zero-bias conductance\/} since
according to Eq.\ (\ref{eq:9}), $\langle\dot X\rangle=LJ$.

As a consequence, for $F=0$ there are always one or more solutions to
Eqs.\ (\ref{eq:10}) and (\ref{eq:11}) with $S_m=0$ and one of these is
the stable one in the ``disordered'' phase.  As argued in Ref.\
\cite{[4]}, for $N\to\infty$ a noise-induced nonequilibrium phase
transition takes place {\em generically\/} towards an ``ordered'' phase
where $P^{st}(-x)\neq P^{st}(x)$. In the present scheme this asymmetry
should be evidenced by the fact that the solution with $S_m=0$ becomes
unstable in favor of two other solutions such that
$P^{st}_{2}(x)=P^{st}_{1}(-x)$, characterised by {\em nonzero\/} values
$\pm|S_m|$.  This fact confers on $S_m$ the rank of an order parameter.

\textbf{The phase boundary:} Since $\sin x$ is an antisymmetric
function, Eq.\ (\ref{eq:11}) results impractical for the task of
finding the curve that separates the ordered phase from the
disordered one, given that on that curve $S_m$ is still zero. For
that goal (exclusively) we solve, instead of Eqs.\ (\ref{eq:10})
and (\ref{eq:11}), the following system:
\begin{equation}
\int_{-L/2}^{L/2}dx\,\cos x\,P^{st}(x,C_m,0)=C_m\mbox{ , }
\int_{-L/2}^{L/2}dx\,\sin x\,\left.\frac{\partial P^{st}}
{\partial_{S_m}}\right|_{S_m=0}=1.\label{eq:12}
\end{equation}

\section{Numerical results}\label{sec.4}
Figure \ref{fig:1} displays (in the same scale as in Ref.\ \cite{[6]})
the phase diagram obtained by solving Eqs.\ (\ref{eq:12}) by the
Newton-Raphson method.  In the region enclosed by the thick lines
(``ordered region'') the {\em stable\/} solution to Eqs.\ (\ref{eq:10}) and
(\ref{eq:11}) has $S_m\neq 0$.  For $\tau$ not too large this
noise-induced phase transition is {\em reentrant\/} as a function of
$Q$, for $K_0=\mathrm{const}$ (a fact already known for $\tau=0$
\cite{[4],[6]}).  The novelty is that for {\em any\/} $\tau\neq 0$, the
phase transition is also reentrant as a function of $K_0$ for
$Q=\mathrm{const}$ (a feature found in Ref.\ \cite{[5]} for a
different system).

The multiplicity of mean-field solutions in the ordered region,
together with the fact that some of them may suddenly disappear as
either $K_0$ or $Q$ are varied (a fact that is closely related to the
occurrence of anomalous hysteresis) hinder picking out the right
solution in this region.  A more systematic characterization of the
aforementioned multiple solutions is achieved when the branch to which
they belong is traced from its corresponding ``homogeneous'' ($S_m=0$)
solution. Accordingly, the thin lines in Fig.\ \ref{fig:1} separate two
sectors within the ordered region with regard to the {\em homogeneous\/}
solutions. Below them (``noise-driven regime'' or NDR) there is {\em a
single\/} solution with $S_m=0$ and $C_m<0$ (as already suggested, in
this regime a solution with $C_m<0$ {\em can\/} be stable since it
corresponds to shaking violently the pendula). Above them
(``interaction-driven regime'' or IDR) there are {\em three\/}
solutions: two of them have opposite signs and (for $K_0/Q$ large
enough) $|C_m|\simeq 0.9$; the remaining one has $C_m\approx 0$. Note
that this line presents a cusp whose meaning was discussed in Ref.\
\cite{[6]}, in  relation with the character of the hysteresis loop.  We
have studied the shape of $P^{st}(x)$ and the behaviour of
$\langle\dot X\rangle$ as a function of $F$ for different locations in
this $(Q,K_0)$ diagram.  The square in Fig.\ \ref{fig:1} indicates a
position inside the ordered zone for which the shape of the SPDF and
the hysteresis cycle are followed as functions of $\tau$. This point
lies in the NDR for $\tau=0$ and in the IDR for $\tau\neq 0$
(marginally so for $\tau=.1$).

Figure \ref{fig:2} shows (for the {\em true\/} solution, namely the
{\em stable\/} $S_m\neq 0$ one) the evolution of $P^{st}(x)$ as a
function of $\tau$, for the state indicated by the square in Fig.\
\ref{fig:1} ($Q=10$, $K_0=10.2$).  The SPDF is always an
{\em asymmetric\/} function of $x$, indicating a {\em spontaneous\/}
breakdown of parity (since $V$ and $W$ remain symmetric): the system
has to choose between two possible {\em asymmetric\/} solutions, of
which just one is shown.  As discussed in Ref.\ \cite{[6]},
$P^{st}(x)$ is {\em bimodal\/} in the NDR. As $\tau$ increases, it
becomes {\em unimodal\/} (this is similar to what one achieves by
{\em decreasing\/} $Q$ at $\tau=0$).

Figures \ref{fig:3}(a) to \ref{fig:3}(c) present, for the state
indicated with a square in Fig.\ \ref{fig:1}, a sequence of
$\langle\dot X\rangle$ vs $F$ plots obtained as $\tau$ increases.  All
the solutions to Eqs.\ (\ref{eq:10}) and (\ref{eq:11}), except the one
belonging to the branch starting at $C_m\approx 0$ for $S_m=0$, have
been included in these figures.  The sequence is analogous to the one
depicted in Fig.\ 9 of Ref.\ \cite{[6]}.  In both cases we see a
crossing from the NDR to the IDR (increasing $\tau$ at fixed $K_0$ has
similar effects as increasing $K_0$ at $\tau=0$, for fixed $Q$).

\section{Conclusions}\label{sec.5}
From the analysis of the foregoing results, we draw the following
conclusions:
\begin{enumerate}
    \item The increase of $\tau$ tends to destroy order, and between
    $\tau=0.1$ and $\tau=0.5$ there exists a new reentrance with respect
    to $K_0$.  In this range of $\tau$, the existence region of the
    ordered state is strongly shrunk.  The qualitative similarity with
    the result arrived at in Ref.\ \cite{[5]}\textemdash in spite of the
    fact that both systems are different\textemdash shows the robustness
    of this result.

    \item Regarding the $\langle\dot X\rangle$ vs $F$ plots one sees that
    as $\tau$ increases the hysteresis cycle becomes more complex, and
    the range of values of $\langle\dot X\rangle$ corresponding to $F\ne
    0$ is severely limited.
\end{enumerate}
Although all of our results stem from a MFA, we see that this
approximation is able to reveal the richness of the phase diagram of
this model.  Moreover, the mean-field results coincide with the
numerical simulations we are undertaking and that will be published
elsewhere \cite{[8]}.

\textbf{Acknowledgments:} The authors thank V. Grunfeld for a critical
revision of the manuscript.  This work was partially supported by the
Argentine Research Council (CONICET) through grant PIP 4953/97.

\newpage
\begin{figure}
\caption{Phase diagram of the model for $T=2.0$, $A=0.15$, and
$F=0.0$. Full lines: $\tau=0.0$; dashed lines: $\tau=0.1$; dotted
lines: $\tau=0.3$.  For each value of $\tau$, the ordered region
lies above and to the right of the corresponding thick line.
Above the thin lines there may exist several solutions when
$S_m\neq 0$, whereas below them there may exist at most one.  The
square corresponds to $Q=10.0$, $K_0=10.2$.} \label{fig:1}
\end{figure}

\begin{figure}
\caption{Shape of $P^{st}(x)$ at the point marked with a square in
Fig.\ 1 ($Q=10.0$, $K_0=10.2$), for $\tau=$0.0 (full line), 0.1
(dashed line), and 0.3 (dotted line).  The other parameters as in
Fig.\ 1.} \label{fig:2}
\end{figure}

\begin{figure}
\caption{The order parameter $V_m=\langle\dot X\rangle$ (particle
current) as a function of $F$ for $Q=10.0$ and $K_0=10.2$ (the square
in Fig.\ 1).  The sequence illustrates the change in the character of
the hysteresis cycle as $\tau$ varies: (a) $\tau=$0.0, (b) $\tau=$0.1,
(c) $\tau=$0.3.  Remaining parameters as in Fig.\ 1.}
\label{fig:3}
\end{figure}

\begin{thebibliography}{99}
\bibitem{[1]} R. D. Vale and F. Oosawa, Adv.\ Biophys.\ {\bf 26}, 97
(1990); A. Ajdari and J. Prost, C. R. Acad.\ Sci., Ser.\ II: Mec.,
Phys., Chim., Sci.\ Terre Univers.\ {\bf 315}, 1635 (1992).

\bibitem{[2]} M. O. Magnasco, Phys.\ Rev.\ Lett.\ {\bf 71}, 1477
(1993); R. D. Astumian and M. Bier, {\em ibid\/} {\bf 72}, 1766 (1994);
C. R. Doering, W. Horsthemke, and J. Riordan, {\em ibid\/} {\bf 72},
2984 (1994); R. Bartussek, P. H\"anggi and J. G. Kissner, Europhys.\
Lett.\ {\bf 28}, 459 (1994).

\bibitem{[3]} P. Reimann, Phys.\ Rep.\ {\bf 290}, 149 (1997).

\bibitem{[4]} P. Reimann, R. Kawai, C. Van den Broeck, and P. H\"anggi,
Europhys.\ Lett.\ {\bf 45}, 545 (1999).

\bibitem{[5]} S. Mangioni, R. Deza, H. S. Wio, and R. Toral, Phys.\
Rev.\ Lett.\ {\bf 79}, 2389 (1997); S. Mangioni, R. Deza, R. Toral,
and H. S. Wio, Phys.\ Rev.\ E {\bf 61}, 223 (2000).

\bibitem{[6]} S. Mangioni, R. Deza, and H. S. Wio, Phys.\ Rev.\ E {\bf
63}, 041115 (2001).

\bibitem{[7]} J. Buceta, J. M. R. Parrondo, C. Van den Broeck, and J.
de la Rubia, Phys.\ Rev.\ E {\bf 61}, 6287 (2000).

\bibitem{[8]} S. Mangioni, R. Deza, and H. S. Wio, to be submitted to
Phys.\ Rev.\ E.
\end{thebibliography}
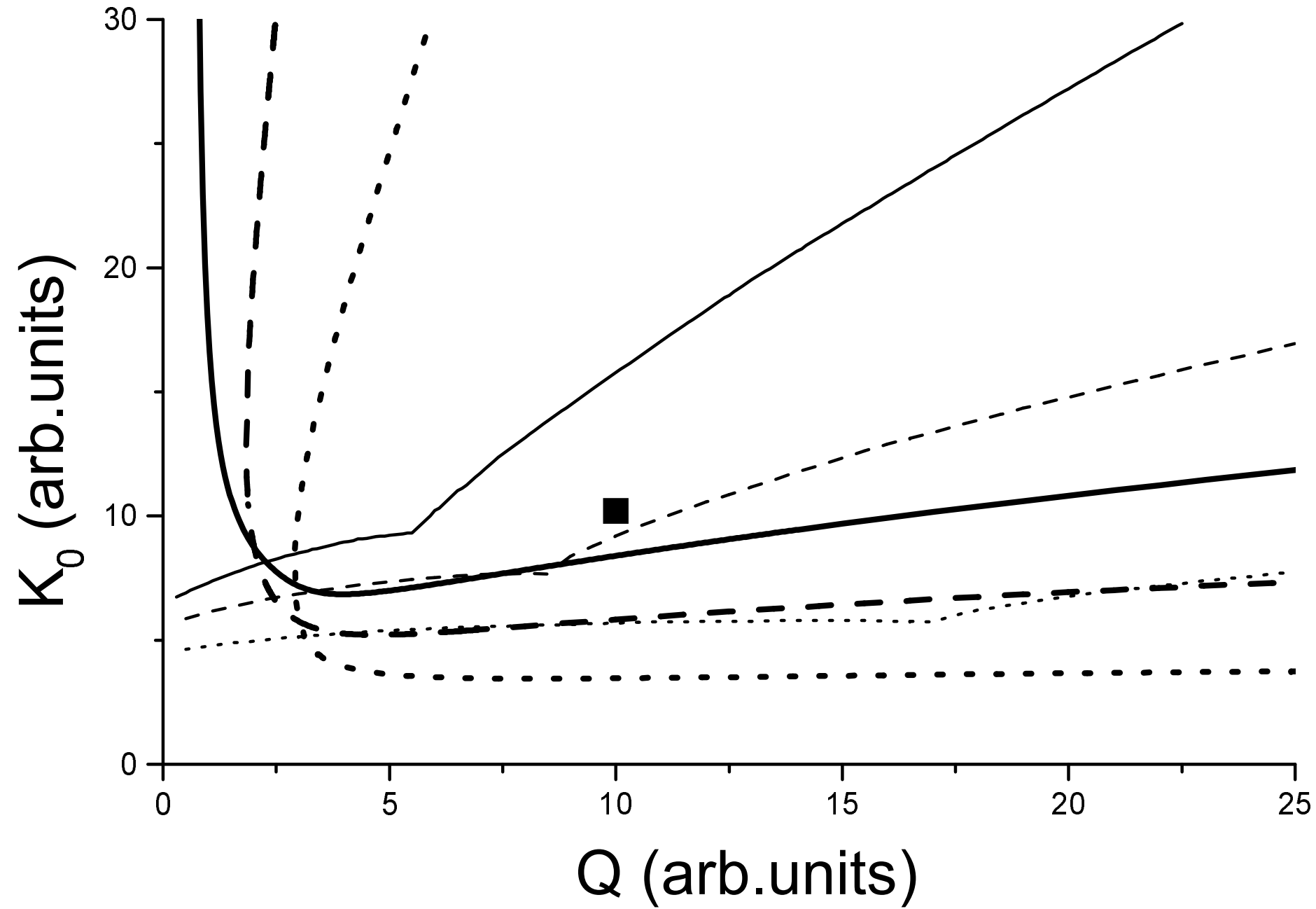
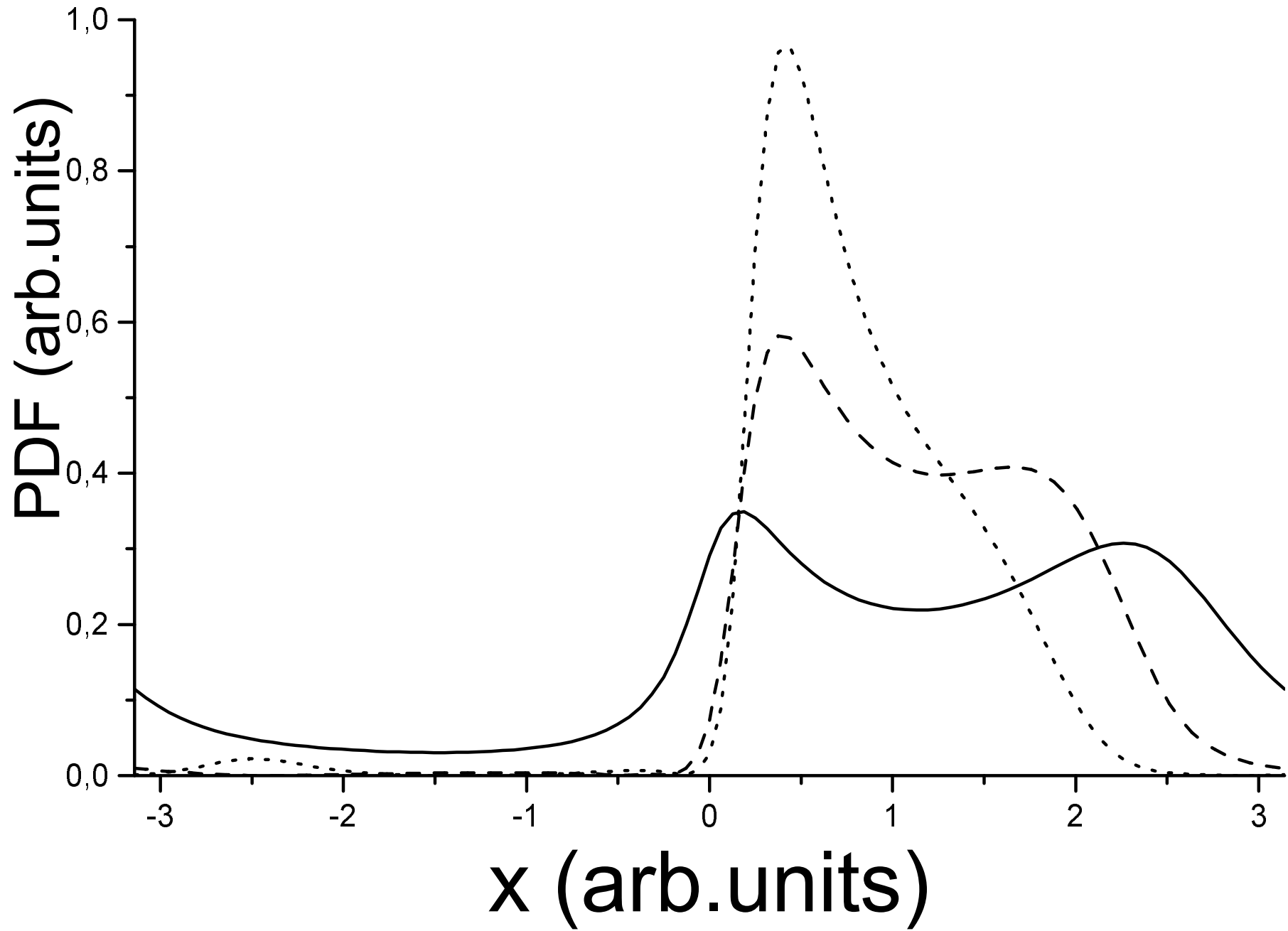
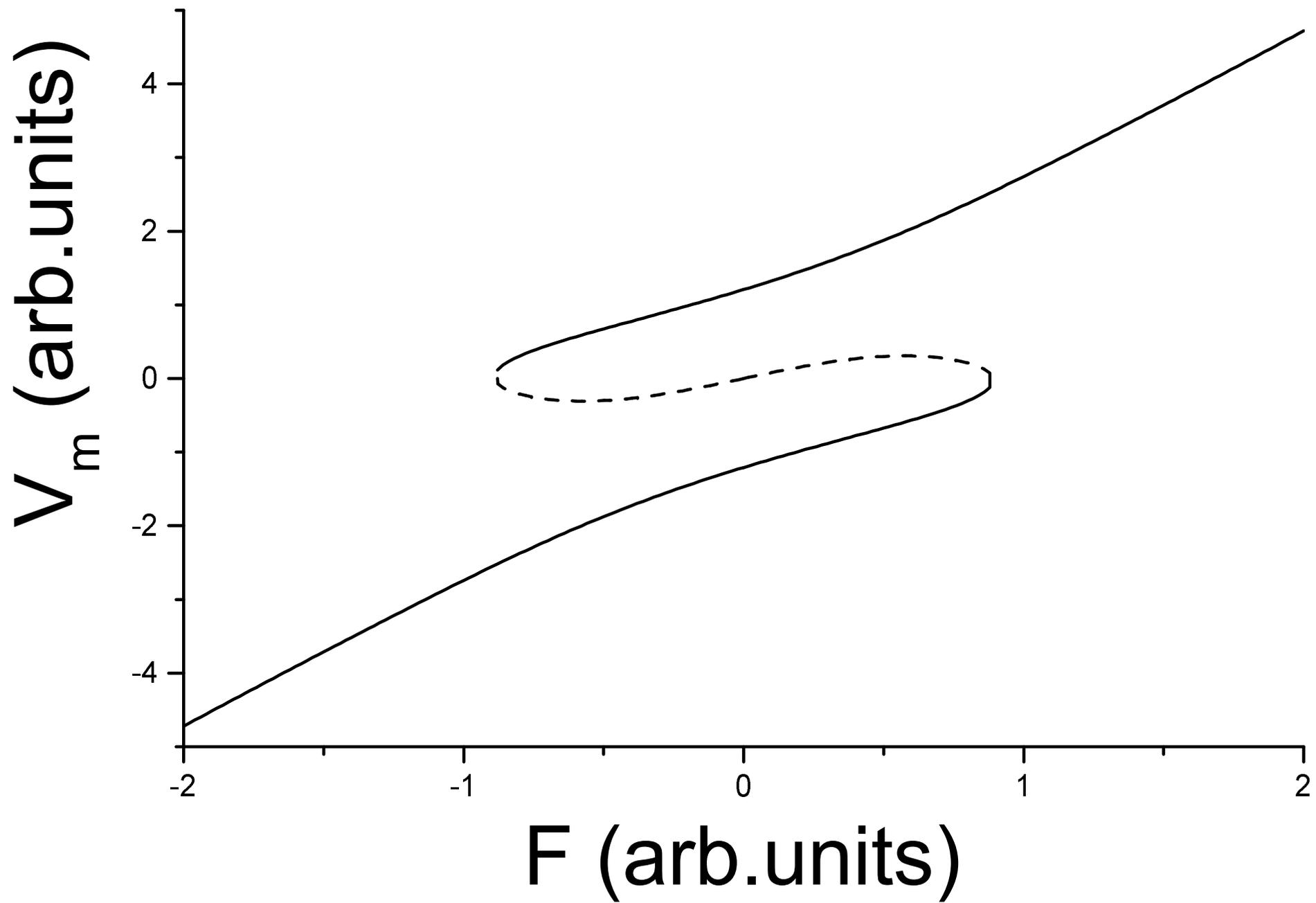
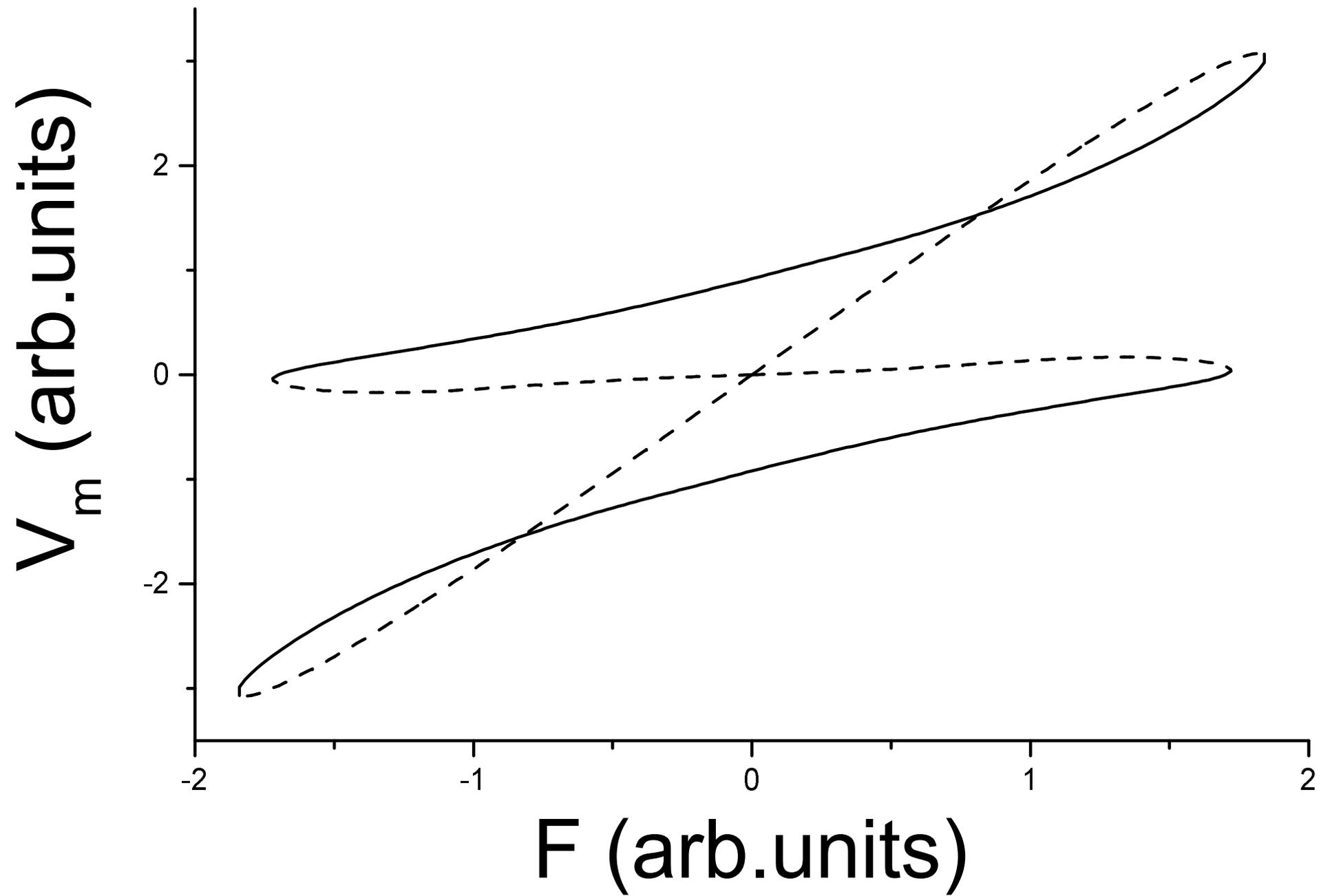
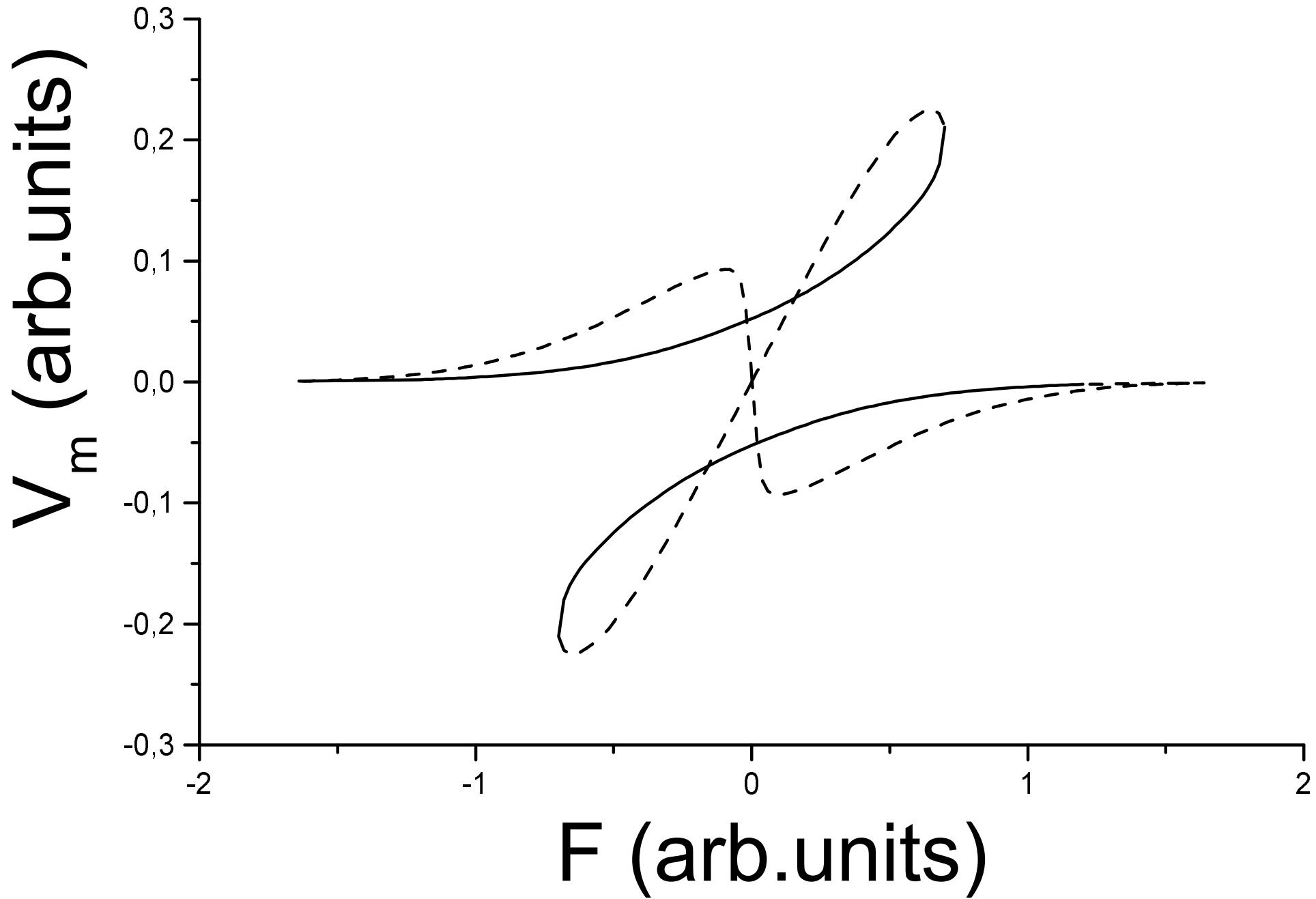
\end{document}